%Paper: astro-ph/9407042
%From: stebbins@perseus.fnal.gov (Albert Stebbins)
%Date: Thu, 14 Jul 94 15:24:27 -0500

% AMS-TEX VERSION 0.999999 - FOR USE WITH TEX VERSION  1.0
% COPYRIGHT (C) 1983 BY AMERICAN MATHEMATICAL SOCIETY

%AJS3
% This is the OLD version of AMSTEX
% It allows the use of roman characters in equations unlike the new version.

% NOTE 1, NOTE 2, ... REFER TO NOTES IN THE FILE AMSTEX.DOC

% SPECIAL CATCODES

\catcode`\@=13                                                         % NOTE 1
\def@{\errmessage{AmS-TeX error: \string@ has no current use
     (use \string\@\space for printed \string@ symbol)}}
\catcode`\@=11                                                         % NOTE 2
\def\@{\char'100 }
\catcode`\~=13                                                         % NOTE 3

% AMSTEX ERROR MESSAGES

\def\err@AmS#1{\errmessage{AmS-TeX error: #1}}                         % NOTE 4

% SOME BASIC CONTROL SEQUENCES USED IN OTHER DEFINITIONS

\def\eat@AmS#1{}

\long\def\comp@AmS#1#2{\def\@AmS{#1}\def\@@AmS{#2}\ifx
   \@AmS\@@AmS\def\cresult@AmS{T}\else\def\cresult@AmS{F}\fi}          % NOTE 5

\def\in@AmS#1#2{\def\intest@AmS##1#1##2{\comp@AmS##2\end@AmS\if T\cresult@AmS
   \def\cresult@AmS{F}\def\in@@AmS{}\else
   \def\cresult@AmS{T}\def\in@@AmS####1\end@AmS{}\fi\in@@AmS}%
   \def\cresult@AmS{F}\intest@AmS#2#1\end@AmS}                         % NOTE 6

% BASIC MECHANICSMS TO ALLOW USER TO MAKE DEFINITIONS

\let\relax@AmS=\relax                                                  % NOTE 7

% CHANGES IN plain WHERE THERE IS \relax, THAT MUST NOW BE \relax@AmS

\def\magstep#1{\ifcase#1 \@m\or 1200\or 1440\or 1728\or 2074\or 2488\fi
     \relax@AmS}

\def\iterate{\body\let\next\iterate \else\let\next\relax@AmS\fi \next}

\def\enskip{\hskip.5em\relax@AmS}

\def\strut{\relax@AmS\ifmmode\copy\strutbox\else\unhcopy\strutbox\fi}

\let\+=\relax@AmS
\def\sett@b{\ifx\next\+\let\next\relax@AmS
    \def\next{\afterassignment\s@tt@b\let\next}%
  \else\let\next\s@tcols\fi\next}
\def\s@tt@b{\let\next\relax@AmS\us@false\m@ketabbox}

\def\smash{\relax@AmS % \relax@AmS, in case this comes first in \halign
  \ifmmode\def\next{\mathpalette\mathsm@sh}\else\let\next\makesm@sh
  \fi\next}

% (END OF CHANGES TO plain)

\def\define#1{\expandafter\ifx\csname\expandafter\eat@AmS\string#1\endcsname
   \relax@AmS\def\dresult@AmS{\def#1}\else
   \err@AmS{\string#1\space is already defined}\def
      \dresult@AmS{\def\garbage@AmS}\fi\dresult@AmS}                   % NOTE 8

\def\predefine#1#2{\let#1=#2}

% MACROS FOR DEFICIENT KEYBOARDS

\chardef\plus=`+
\chardef\equal=`=
\chardef\less=`<
\chardef\more=`>

% MACROS FOR HANDLING TEXT

\let\ic@AmS=\/
\def\/{\unskip\ic@AmS}

\def\Space@AmS.{\futurelet\Space@AmS\relax@AmS}
\Space@AmS. %                                             % NOTE 10 (no NOTE 9)

\def~{\unskip\futurelet\tok@AmS\s@AmS}
\def\s@AmS{\ifx\tok@AmS\Space@AmS\def\next@AmS{}\else
        \def\next@AmS{\ }\fi\penalty 9999 \next@AmS}                  % NOTE 11

\def\period{\unskip.\spacefactor3000 { }}

\def\srdr@AmS{\thinspace}                                             % NOTE 12
\def\drsr@AmS{\kern .02778em }
\def\sldl@AmS{\kern .02778em}
\def\dlsl@AmS{\thinspace}

\def\lqtest@AmS#1{\comp@AmS{#1}`\if T\cresult@AmS\else\comp@AmS{#1}\lq\fi}

                                                                      % NOTE 13

\def\qspace#1{\unskip
  \lqtest@AmS{#1}\let\fresult@AmS=\cresult@AmS\if T\cresult@AmS
     \def\qspace@AmS{\ifx\tok@AmS\Space@AmS\def\next@AmS{\dlsl@AmS`}\else
       \def\next@AmS{\qspace@@AmS}\fi\next@AmS}\else
     \def\qspace@AmS{\ifx\tok@AmS\Space@AmS\def\next@AmS{\drsr@AmS'}\else
       \def\next@AmS{\qspace@@AmS}\fi\next@AmS}\fi
    \futurelet\tok@AmS\qspace@AmS}                                    % NOTE 14

\def\qspace@@AmS{\futurelet\tok@AmS\qspace@@@AmS}

\def\qspace@@@AmS{\if T\fresult@AmS  \ifx\tok@AmS`\sldl@AmS`\else
       \ifx\tok@AmS\lq\sldl@AmS`\else \dlsl@AmS`\fi \fi
                         \else  \ifx\tok@AmS'\srdr@AmS'\else
        \ifx\tok@AmS\rq\srdr@AmS'\else \drsr@AmS'\fi \fi
        \fi}

\def\{{\relax@AmS\ifmmode\delimiter"4266308 \else
    $\delimiter"4266308 $\fi}                            % NOTE 16 (No NOTE 15)

\def\}{\relax@AmS\ifmmode\delimiter"5267309 \else$\delimiter"5267309 $\fi}

\def\AmSTeX{$\cal A$\kern-.1667em\lower.5ex\hbox{$\cal M$}\kern-.125em
     $\cal S$-\TeX}

                            % NOTE 17

\def\linebreak{\unskip\penalty-10000 }                                % NOTE 18
\def\pagebreak{\vadjust{\penalty-10000 }}

\def\newline{\ifvmode \err@AmS{There's no line here to break}\else
     \hfil\penalty-10000 \fi}

\def\topspace#1{\insert\topins{\penalty100 \splittopskip=0pt
     \vbox to #1{}}}
\def\midspace#1{\setbox0=\vbox to #1{}\advance\dimen0 by \pagetotal
  \ifdim\dimen0>\pagegoal\topspace{#1}\else\vadjust{\box0}\fi}

\long\def\comment{\begingroup
 \catcode`\{=12 \catcode`\}=12 \catcode`\#= 12 \catcode`\^^M=12
   \catcode`\%=12 \catcode`^^A=14
    \comment@AmS}
\begingroup\catcode`^^A=14
\catcode`\^^M=12  ^^A
\long\gdef\comment@AmS#1^^M#2{\comp@AmS\endcomment{#2}\if T\cresult@AmS^^A
\def\comment@@AmS{\endgroup}\else^^A
 \long\def\comment@@AmS{\comment@AmS#2}\fi\comment@@AmS}\endgroup     % NOTE 19

% STYLE, SPACING AND ALTERNATE NAMES

\def\text#1{\hbox{\rm#1}}

\def\quad{\relax@AmS\ifmmode
    \hbox{\hskip1em}\else\hskip1em\relax@AmS\fi}                      % NOTE 20
\def\qquad{\quad\quad}
\def\,{\relax@AmS\ifmmode\mskip\thinmuskip\else$\mskip\thinmuskip$\fi}
\def\;{\relax@AmS
  \ifmmode\mskip\thickmuskip\else$\mskip\thickmuskip$\fi}

\def\frac#1#2{{#1\over#2}}

\mathchardef\:="603A                                                  % NOTE 21

% BIG DELIMITERS

\def\big@AmS#1{{\hbox{$\left#1\vbox to\big@@AmS{}\right.\offspace@AmS$}}}
\def\Big@AmS#1{{\hbox{$\left#1\vbox to\Big@@AmS{}\right.\offspace@AmS$}}}
\def\bigg@AmS#1{{\hbox{$\left#1\vbox to\bigg@@AmS{}\right.\offspace@AmS$}}}
\def\Bigg@AmS#1{{\hbox{$\left#1\vbox to\Bigg@@AmS{}\right.\offspace@AmS$}}}
\def\offspace@AmS{\nulldelimiterspace0pt \mathsurround0pt }

\def\big@@AmS{8.5pt}                                % NOTE 24 (no NOTES 22, 23)
\def\Big@@AmS{11.5pt}
\def\bigg@@AmS{14.5pt}
\def\Bigg@@AmS{17.5pt}

\def\bigl{\mathopen\big@AmS}
\def\bigm{\mathrel\big@AmS}
\def\bigr{\mathclose\big@AmS}
\def\Bigl{\mathopen\Big@AmS}
\def\Bigm{\mathrel\Big@AmS}
\def\Bigr{\mathclose\Big@AmS}
\def\biggl{\mathopen\bigg@AmS}
\def\biggm{\mathrel\bigg@AMS}
\def\biggr{\mathclose\bigg@AmS}
\def\Biggl{\mathopen\Bigg@AmS}
\def\Biggm{\mathrel\Bigg@AmS}
\def\Biggr{\mathclose\Bigg@AmS}

%  MAKING ' WORK FOR PRIMES

{\catcode`'=13 \gdef'{^\bgroup\prime\prime@AmS}}
\def\prime@AmS{\futurelet\tok@AmS\prime@@AmS}
\def\prime@@@AmS#1{\futurelet\tok@AmS\prime@@AmS}
\def\prime@@AmS{\ifx\tok@AmS'\def\next@AmS{\prime\prime@@@AmS}\else
   \def\next@AmS{\egroup}\fi\next@AmS}

%  SMASHES                                                            % NOTE 25

\def\topsmash{\relax@AmS\ifmmode\def\topsmash@AmS
   {\mathpalette\mathtopsmash@AmS}\else
    \let\topsmash@AmS=\maketopsmash@AmS\fi\topsmash@AmS}

\def\maketopsmash@AmS#1{\setbox0=\hbox{#1}\topsmash@@AmS}

\def\mathtopsmash@AmS#1#2{\setbox0=\hbox{$#1{#2}$}\topsmash@@AmS}

\def\topsmash@@AmS{\vbox to 0pt{\kern-\ht0\box0}}

\def\botsmash{\relax@AmS\ifmmode\def\botsmash@AmS
   {\mathpalette\mathbotsmash@AmS}\else
     \let\botsmash@AmS=\makebotsmash@AmS\fi\botsmash@AmS}

\def\makebotsmash@AmS#1{\setbox0=\hbox{#1}\botsmash@@AmS}

\def\mathbotsmash@AmS#1#2{\setbox0=\hbox{$#1{#2}$}\botsmash@@AmS}

\def\botsmash@@AmS{\vbox to \ht0{\box0\vss}}

%  LARGE OPERATORS

\def\LimitsOnSums{\let\slimits@AmS=\displaylimits}                    % NOTE 26
\def\NoLimitsOnSums{\let\slimits@AmS=\nolimits}

\LimitsOnSums

\mathchardef\coprod@AmS"1360       \def\coprod{\coprod@AmS\slimits@AmS}
\mathchardef\bigvee@AmS"1357       \def\bigvee{\bigvee@AmS\slimits@AmS}
\mathchardef\bigwedge@AmS"1356     \def\bigwedge{\bigwedge@AmS\slimits@AmS}
\mathchardef\biguplus@AmS"1355     \def\biguplus{\biguplus@AmS\slimits@AmS}
\mathchardef\bigcap@AmS"1354       \def\bigcap{\bigcap@AmS\slimits@AmS}
\mathchardef\bigcup@AmS"1353       \def\bigcup{\bigcup@AmS\slimits@AmS}
\mathchardef\prod@AmS"1351         \def\prod{\prod@AmS\slimits@AmS}
\mathchardef\sum@AmS"1350          \def\sum{\sum@AmS\slimits@AmS}
\mathchardef\bigotimes@AmS"134E    \def\bigotimes{\bigotimes@AmS\slimits@AmS}
\mathchardef\bigoplus@AmS"134C     \def\bigoplus{\bigoplus@AmS\slimits@AmS}
\mathchardef\bigodot@AmS"134A      \def\bigodot{\bigodot@AmS\slimits@AmS}
\mathchardef\bigsqcup@AmS"1346     \def\bigsqcup{\bigsqcup@AmS\slimits@AmS}

\def\LimitsOnInts{\let\ilimits@AmS=\displaylimits}
\def\NoLimitsOnInts{\let\ilimits@AmS=\nolimits}

\NoLimitsOnInts

\mathchardef\int@AmS"1352
\def\int{\gdef\intflag@AmS{T}\int@AmS\ilimits@AmS}                    % NOTE 27

\mathchardef\oint@AmS"1348 \def\oint{\gdef\intflag@AmS{T}\oint@AmS\ilimits@AmS}

\def\inttest@AmS#1{\def\intflag@AmS{F}\setbox0=\hbox{$#1$}}

\def\intic@AmS{\mathchoice{\hbox{\hskip5pt}}{\hbox
          {\hskip4pt}}{\hbox{\hskip4pt}}{\hbox{\hskip4pt}}}           % NOTE 28
\def\negintic@AmS{\mathchoice
  {\hbox{\hskip-5pt}}{\hbox{\hskip-4pt}}{\hbox{\hskip-4pt}}{\hbox{\hskip-4pt}}}
\def\intkern@AmS{\mathchoice{\!\!\!}{\!\!}{\!\!}{\!\!}}
\def\intdots@AmS{\mathchoice{\cdots}{{\cdotp}\mkern 1.5mu
    {\cdotp}\mkern 1.5mu{\cdotp}}{{\cdotp}\mkern 1mu{\cdotp}\mkern 1mu
      {\cdotp}}{{\cdotp}\mkern 1mu{\cdotp}\mkern 1mu{\cdotp}}}

\newcount\intno@AmS                                                   % NOTE 29

\def\intii{\gdef\intflag@AmS{T}\intno@AmS=2\futurelet                 % NOTE 30
              \tok@AmS\ints@AmS}
\def\intiii{\gdef\intflag@AmS{T}\intno@AmS=3\futurelet\tok@AmS\ints@AmS}
\def\intiv{\gdef\intflag@AmS{T}\intno@AmS=4\futurelet\tok@AmS\ints@AmS}
\def\intdotsint{\gdef\intflag@AmS{T}\intno@AmS=0\futurelet
    \tok@AmS\ints@AmS}

\def\ints@AmS{\findlimits@AmS\ints@@AmS}

\def\findlimits@AmS{\def\ignoretoken@AmS{T}\ifx\tok@AmS\limits
   \def\limits@AmS{T}\else\ifx\tok@AmS\nolimits\def\limits@AmS{F}\else
     \def\ignoretoken@AmS{F}\ifx\ilimits@AmS\nolimits\def\limits@AmS{F}\else
       \def\limits@AmS{T}\fi\fi\fi}

\def\multintlimits@AmS{\int@AmS\ifnum \intno@AmS=0\intdots@AmS
  \else \intkern@AmS\fi
    \ifnum\intno@AmS>2\int@AmS\intkern@AmS\fi
     \ifnum\intno@AmS>3 \int@AmS\intkern@AmS\fi \int@AmS}

\def\multint@AmS{\int\ifnum \intno@AmS=0\intdots@AmS\else\intkern@AmS\fi
   \ifnum\intno@AmS>2\int\intkern@AmS\fi
    \ifnum\intno@AmS>3 \int\intkern@AmS\fi \int}

\def\ints@@AmS{\if F\ignoretoken@AmS\def\ints@@@AmS{\if
    T\limits@AmS\negintic@AmS
 \mathop{\intic@AmS\multintlimits@AmS}\limits\else
    \multint@AmS\nolimits\fi}\else\def\ints@@@AmS{\if T\limits@AmS
   \negintic@AmS\mathop{\intic@AmS\multintlimits@AmS}\limits\else
    \multint@AmS\nolimits\fi\eat@AmS}\fi\ints@@@AmS}

\def\LimitsOnNames{\let\nlimits@AmS=\displaylimits}
\def\NoLimitsOnNames{\let\nlimits@AmS=\nolimits}

\LimitsOnNames

                                              % NOTE 31

\def\operatornamewithlimits#1{\mathop{\mathcode`'="7027 \mathcode`-="702D
   \rm #1}\nlimits@AmS}

\def\liminj{\setbox0=\hbox{\rm lim}\mathop{\rm lim}
		\limits_{\topsmash{\hbox to \wd0{\leftarrowfill}}}}
\def\limproj{\setbox0=\hbox{\rm lim}\mathop{\rm lim}
		\limits_{\topsmash{\hbox to \wd0{\rightarrowfill}}}}

% SUBSIDIARY CONSIDERATIONS FOR LARGE OPERATORS -- BUFFER AND SHAVE

\newdimen\buffer@AmS
\buffer@AmS=\fontdimen13\tenex                                        % NOTE 32
\newdimen\buffer
\buffer=\buffer@AmS

\def\resetbuffer{\fontdimen13 \tenex=\buffer@AmS \buffer=\buffer@AmS}

% ALIGNED UNITS

\def\Let@AmS{\relax@AmS\iffalse{\fi\let\\=\cr\iffalse}\fi}            % NOTE 34

\def\align{\def\vspace##1{\noalign{\vskip ##1}}                       % NOTE 35
  \,\vcenter\bgroup\Let@AmS\tabskip=0pt\openup3pt\mathsurround=0pt  % NOTE 35.1
  \halign\bgroup\strut
  \hfil$\displaystyle{##}$&$\displaystyle{{}##}$\hfil\cr}        % NOTES 36, 37

\def\endalign{\strut\crcr\egroup\egroup}

\def\bunch{\def\vspace##1{\noalign{\vskip ##1}}
  \,\vcenter\bgroup\Let@AmS\tabskip=0pt\openup3pt\mathsurround=0pt
     \halign\bgroup\strut\hfil$\displaystyle{##}$\hfil\cr}

\def\endbunch{\strut\crcr\egroup\egroup}

\def\matrix{\catcode`\^^I=4 \futurelet\tok@AmS\matrix@AmS}            % NOTE 38

\def\matrix@AmS{\relax@AmS\iffalse{\fi \ifnum`}=0\fi\ifx\tok@AmS\format
   \def\next@AmS{\expandafter\matrix@@AmS\eat@AmS}\else
   \def\next@AmS{\matrix@@@AmS}\fi\next@AmS}

\def\matrix@@@AmS{
 \ifnum`{=0\fi\iffalse}\fi\,\vcenter\bgroup\Let@AmS\tabskip=0pt
    \normalbaselines\halign\bgroup $\strut\hfil##\hfil$&&\quad$\strut
  \hfil##\hfil$\cr\strut\cr\noalign{\kern-\baselineskip}}             % NOTE 39

\def\matrix@@AmS#1\\{
   \def\premable@AmS{#1}\toks@{##}
 \def\c{$\copy\strutbox\hfil\the\toks@\hfil$}\def\r
   {$\copy\strutbox\hfil\the\toks@$}%
   \def\l{$\copy\strutbox\the\toks@\hfil$}%
\setbox0=
\hbox{\xdef\Preamble@AmS{\premable@AmS}}
 \def\vspace##1{\noalign{\vskip ##1}}\ifnum`{=0\fi\iffalse}\fi
\,\vcenter\bgroup\Let@AmS
  \tabskip=0pt\normalbaselines\halign\bgroup\span\Preamble@AmS\cr
    \mathstrut\cr\noalign{\kern-\baselineskip}}
                                                                      % NOTE 40

\def\endmatrix{\crcr\mathstrut\cr\noalign{\kern-\baselineskip
   }\egroup\egroup\,\catcode`\^^I=10 }

\def\spacedots#1for#2{\multispan#2\leaders\hbox{$\mkern#1mu.\mkern
    #1mu$}\hfill}
                                 % NOTE 41

\def\enabletabs{\catcode`\^^I=4 \enabletabs@AmS}
\def\enabletabs@AmS#1\disabletabs{#1\catcode`\^^I=10 }                % NOTE 42

\def\smallmatrix{\futurelet\tok@AmS\smallmatrix@AmS}                  % NOTE 43

\def\smallmatrix@AmS{\relax@AmS\iffalse{\fi \ifnum`}=0\fi\ifx\tok@AmS\format
   \def\next@AmS{\expandafter\smallmatrix@@AmS\eat@AmS}\else
   \def\next@AmS{\smallmatrix@@@AmS}\fi\next@AmS}

\def\smallmatrix@@@AmS{
 \ifnum`{=0\fi\iffalse}\fi\,\vcenter\bgroup\Let@AmS\tabskip=0pt
    \baselineskip8pt\lineskip1pt\lineskiplimit1pt
  \halign\bgroup $\strut\hfil##\hfil$&&\;$\strut
  \hfil##\hfil$\cr\strut\cr\noalign{\kern-\baselineskip}}

\def\smallmatrix@@AmS#1\\{
   \def\premable@AmS{#1}\toks@{##}
 \def\c{$\copy\strutbox\hfil\the\toks@\hfil$}\def\r
   {$\copy\strutbox\hfil\the\toks@$}%
   \def\l{$\copy\strutbox\the\toks@\hfil$}%
\hbox{\xdef\Preamble@AmS{\premable@AmS}}
 \def\vspace##1{\noalign{\vskip ##1}}\ifnum`{=0\fi\iffalse}\fi
\,\vcenter\bgroup\Let@AmS
     \tabskip=0pt\baselineskip8pt\lineskip1pt\lineskiplimit1pt
\halign\bgroup\span\Preamble@AmS\cr
    \mathstrut\cr\noalign{\kern-\baselineskip}}

\def\endsmallmatrix{\crcr\mathstrut\cr\noalign{\kern-\baselineskip}
   \egroup\egroup\,}

\def\cases{\left\{ \,\vcenter\bgroup\Let@AmS\normalbaselines\tabskip=0pt
   \halign\bgroup$##\hfil$&\qquad$##\hfil$\cr}                        % NOTE 44

\def\endcases{\crcr\egroup\egroup\right.}

% TAGGING

\def\TagsOnLeft{\def\tagposition@AmS{L}}
\def\TagsOnRight{\def\tagposition@AmS{R}}
\def\TagsAsMath{\def\tagstyle@AmS{M}}
\def\TagsAsText{\def\tagstyle@AmS{T}}

\TagsOnLeft
\TagsAsText

\def\tag#1$${\if L\tagposition@AmS
    \leqno\else\eqno\fi\def\atag@AmS{T}\maketag@AmS#1\tagend@AmS$$}   % NOTE 45

\def\maketag@AmS{\futurelet\tok@AmS\maketag@@AmS}                     % NOTE 46
\def\maketag@@AmS{\ifx\tok@AmS[\def\next@AmS{\maketag@@@AmS}\else
      \def\next@AmS{\maketag@@@@AmS}\fi\next@AmS}
\def\maketag@@@AmS[#1]#2\tagend@AmS{\if F\atag@AmS\else             % NOTE 46.1
   \if M\tagstyle@AmS\hbox{$#1$}\else\hbox{#1}\fi\fi
       \gdef\atag@AmS{F}}
\def\maketag@@@@AmS#1\tagend@AmS{\if F\atag@AmS \else
        \if T\autotag@AmS \setbox0=\hbox
    {\if M\tagstyle@AmS\tagform@AmS{$#1$}\else\tagform@AmS{#1}\fi}
                        \ifdim\wd0=0pt \tagform@AmS{*}\else
            \if M\tagstyle@AmS\tagform@AmS{$#1$}\else\tagform@AmS{#1}\fi
                     \fi\else
               \if M\tagstyle@AmS\tagform@AmS{$#1$}\else\tagform@AmS{#1}\fi
                     \fi
                  \fi\gdef\atag@AmS{F}}

\def\tagform@AmS#1{\hbox{\rm(#1\unskip)}}

\def\AutoTag{\def\autotag@AmS{T}}
\def\NoAutoTag{\def\autotag@AmS{F}}

\NoAutoTag

\def\inaligntag@AmS{F} \def\inbunchtag@AmS{F}                         % NOTE 47

\def\CenteredTagsOnBrokens{\def\centerbroken@AmS{T}}                  % NOTE 48
\def\TopOrBottomTagsOnBrokens{\def\centerbroken@AmS{F}}
\TopOrBottomTagsOnBrokens

\def\broken{\global\setbox0=\vbox\bgroup\Let@AmS\tabskip=0pt
 \if T\inaligntag@AmS\else
   \if T\inbunchtag@AmS\else\openup3pt\fi\fi\mathsurround=0pt
     \halign\bgroup\strut\hfil$\displaystyle{##}$&$\displaystyle{{}##}$\hfill
      \cr}
                                                                      % NOTE 49
\def\endbroken{\strut\crcr\egroup\egroup
      \global\setbox7=\vbox{\unvbox0\setbox1=\lastbox
      \hbox{\unhbox1\unskip\setbox2=\lastbox
       \unskip\setbox3=\lastbox
         \global\setbox4=\copy3
          \box3\box2}}%                                               % NOTE 50
  \if L\tagposition@AmS
     \if T\inaligntag@AmS
           \if T\centerbroken@AmS\gdef\broken@AmS
                {&\vcenter{\vbox{\moveleft\wd4\box7}}}%               % NOTE 51
           \else
            \gdef\broken@AmS{&\vbox{\moveleft\wd4\vtop{\unvbox7}}}%   % NOTE 52
           \fi
     \else                                                            % NOTE 53
           \if T\centerbroken@AmS\gdef\broken@AmS
                {\vcenter{\box7}}%
           \else
              \gdef\broken@AmS{\vtop{\unvbox7}}%
           \fi
     \fi
  \else                                                  % NOTE 55 (no note 54)
      \if T\inaligntag@AmS
           \if T\centerbroken@AmS
              \gdef\broken@AmS{&\vcenter{\vbox{\moveleft\wd4\box7}}}%
          \else
             \gdef\broken@AmS{&\vbox{\moveleft\wd4\box7}}%
          \fi
      \else
          \if T\centerbroken@AmS
            \gdef\broken@AmS{\vcenter{\box7}}%
          \else
             \gdef\broken@AmS{\box7}%
          \fi
      \fi
  \fi\broken@AmS}

\def\cbroken{\xdef\centerbroken@@AmS{\centerbroken@AmS}%
                       \def\centerbroken@AmS{T}\broken}               % NOTE 56
\def\endcbroken{\endbroken\def\centerbroken@AmS{\centerbroken@@AmS}}

\def\multline#1${\in@AmS\tag{#1}\if T\cresult@AmS
 \def\multline@AmS{\def\atag@AmS{T}\getmltag@AmS#1$}\else
   \def\multline@AmS{\def\atag@AmS{F}\setbox9=\hbox{}\multline@@AmS
    \multline@@@AmS#1$}\fi\multline@AmS}                              % NOTE 57

\def\getmltag@AmS#1\tag#2${\setbox9=\hbox{\maketag@AmS#2\tagend@AmS}%
           \multline@@AmS\multline@@@AmS#1$}

\def\multline@@AmS{\if L\tagposition@AmS
     \def\lwidth@AmS{\hskip\wd9}\def\rwidth@AmS{\hskip0pt}\else
      \def\lwidth@AmS{\hskip0pt}\def\rwidth@AmS{\hskip\wd9}\fi}      % NOTE 58

\def\multline@@@AmS{\def\vspace##1{\noalign{\vskip ##1}}%
 \def\shoveright##1{##1\hfilneg\rwidth@AmS\quad}                      % NOTE 59
  \def\shoveleft##1{\setbox                                           % NOTE 60
      0=\hbox{$\displaystyle{}##1$}%
     \setbox1=\hbox{$\displaystyle##1$}%
     \ifdim\wd0=\wd1
    \hfilneg\lwidth@AmS\quad##1\else
      \setbox2=\hbox{\hskip\wd0\hskip-\wd1}%
       \hfilneg\lwidth@AmS\quad\hskip-.5\wd2 ##1\fi}
     \vbox\bgroup\Let@AmS\openup3pt\halign\bgroup\hbox to \the\displaywidth
      {$\displaystyle\hfil{}##\hfil$}\cr\hfilneg\quad
      \if L\tagposition@AmS\hskip-1em\copy9\quad\else\fi}             % NOTE 61

\def\endmultline{\if R\tagposition@AmS\quad\box9                 % NOTES 62, 63
   \hskip-1em\else\fi\quad\hfilneg\crcr\egroup\egroup}

\def\aligntag#1$${\def\inaligntag@AmS{T}\openup3pt\mathsurround=0pt   % NOTE 64
 \Let@AmS
   \def\tag{\gdef\atag@AmS{T}&}                                       % NOTE 65
   \def\vspace##1{\noalign{\vskip##1}}                                % NOTE 66
    \def\xtext##1{\noalign{\hbox{##1}}}                               % NOTE 67
   \def\break{\noalign{\penalty-10000 }}                              % NOTE 68
   \def\nobreak{\noalign{\penalty 10000 }}
   \def\allowbreak{\noalign{\penalty 0 }}
   \def\goodbreak{\noalign{\penalty -500 }}
    \gdef\atag@AmS{F}%
\if L\tagposition@AmS\laligntag@AmS#1$$\else
   \raligntag@AmS#1$$\fi}

\def\raligntag@AmS#1$${\tabskip\centering
   \halign to \the\displaywidth
{\hfil$\displaystyle{##}$\tabskip 0pt
    &$\displaystyle{{}##}$\hfil\tabskip\centering
   &\llap{\maketag@AmS##\tagend@AmS}\tabskip 0pt\cr\noalign{\vskip-
     \lineskiplimit}#1\crcr}$$}

\def\laligntag@AmS#1$${\tabskip\centering                             % NOTE 69
   \halign to \the\displaywidth
{\hfil$\displaystyle{##}$\tabskip0pt
   &$\displaystyle{{}##}$\hfil\tabskip\centering
    &\kern-\displaywidth\rlap{\maketag@AmS##\tagend@AmS}\tabskip
    \the\displaywidth\cr\noalign{\vskip-\lineskiplimit}#1\crcr}$$}

\def\bunchtag#1$${\def\inbunchtag@AmS{T}\openup3pt\mathsurround=0pt   % NOTE 70
    \Let@AmS
   \def\tag{\gdef\atag@AmS{T}&}
   \def\vspace##1{\noalign{\vskip##1}}
   \def\xtext##1{\noalign{\hbox{##1}}}
   \def\break{\noalign{\penalty-10000 }}
   \def\nobreak{\noalign{\penalty 10000 }}
   \def\allowbreak{\noalign{\penalty 0 }}
    \def\goodbreak{\noalign{\penalty -500 }}
  \if L\tagposition@AmS\lbunchtag@AmS#1$$\else
    \rbunchtag@AmS#1$$\fi}

\def\rbunchtag@AmS#1$${\tabskip\centering
    \halign to \displaywidth {$\hfil\displaystyle{##}\hfil$&
      \llap{\maketag@AmS##\tagend@AmS}\tabskip 0pt\cr\noalign{\vskip-
       \lineskiplimit}#1\crcr}$$}

\def\lbunchtag@AmS#1$${\tabskip\centering
   \halign to \displaywidth
{$\hfil\displaystyle{##}\hfil$&\kern-
    \displaywidth\rlap{\maketag@AmS##\tagend@AmS}\tabskip\the\displaywidth\cr
    \noalign{\vskip-\lineskiplimit}#1\crcr}$$}

%  MISCELLANEOUS

                                           % NOTE 71

% CONTINUED FRACTIONS

\def\numeratorleft#1{#1\hskip 0pt plus 1filll\relax@AmS}
\def\numeratorright#1{\hskip 0pt plus 1filll\relax@AmS#1}
\def\numeratorcenter#1{\hskip 0pt plus 1filll\relax@AmS
      #1\hskip 0pt plus 1filll\relax@AmS}

\def\cfrac@AmS#1,{\def\numerator@AmS{#1}\cfrac@@AmS*}                 % NOTE 72

\def\cfrac@@AmS#1;#2#3\cfend@AmS{\comp@AmS\cfmark@AmS{#2}\if T\cresult@AmS
 \gdef\cfrac@@@AmS
  {\expandafter\eat@AmS\numerator@AmS\strut\over\eat@AmS#1}\else
  \comp@AmS;{#2}\if T\cresult@AmS\gdef\cfrac@@@AmS
  {\expandafter\eat@AmS\numerator@AmS\strut\over\eat@AmS#1}\else
\gdef\cfrac@@@AmS{\if R\cftype@AmS\hfill\else\fi
    \expandafter\eat@AmS\numerator@AmS\strut
    \if L\cftype@AmS\hfill\else\fi\over
       \eat@AmS#1\displaystyle {\cfrac@AmS*#2#3\cfend@AmS}}
     \fi\fi\cfrac@@@AmS}

\def\cfrac#1{\def\cftype@AmS{C}\cfrac@AmS*#1;\cfmark@AmS\cfend@AmS}

\def\cfracl#1{\def\cftype@AmS{L}\cfrac@AmS*#1;\cfmark@AmS\cfend@AmS}

\def\cfracr#1{\def\cftype@AmS{R}\cfrac@AmS*#1;\cfmark@AmS\cfend@AmS}

%  ARROWS                                                             % NOTE 73

\def\overrightarrow{\mathpalette\overrightarrow@AmS}

\def\overrightarrow@AmS#1#2{\vbox{\halign{$##$\cr
    #1{-}\mkern-6mu\cleaders\hbox{$#1\mkern-2mu{-}\mkern-2mu$}\hfill
     \mkern-6mu{\to}\cr
     \noalign{\kern -1pt\nointerlineskip}
     \hfil#1#2\hfil\cr}}}

\def\overleftarrow{\mathpalette\overleftarrow@Ams}

\def\overleftarrow@Ams#1#2{\vbox{\halign{$##$\cr
     #1{\leftarrow}\mkern-6mu\cleaders\hbox{$#1\mkern-2mu{-}\mkern-2mu$}\hfill
      \mkern-6mu{-}\cr
     \noalign{\kern -1pt\nointerlineskip}
     \hfil#1#2\hfil\cr}}}

\def\overleftrightarrow{\mathpalette\overleftrightarrow@AmS}

\def\overleftrightarrow@AmS#1#2{\vbox{\halign{$##$\cr
     #1{\leftarrow}\mkern-6mu\cleaders\hbox{$#1\mkern-2mu{-}\mkern-2mu$}\hfill
       \mkern-6mu{\to}\cr
    \noalign{\kern -1pt\nointerlineskip}
      \hfil#1#2\hfil\cr}}}

\def\underrightarrow{\mathpalette\underrightarrow@AmS}

\def\underrightarrow@AmS#1#2{\vtop{\halign{$##$\cr
    \hfil#1#2\hfil\cr
     \noalign{\kern -1pt\nointerlineskip}
    #1{-}\mkern-6mu\cleaders\hbox{$#1\mkern-2mu{-}\mkern-2mu$}\hfill
     \mkern-6mu{\to}\cr}}}

\def\underleftarrow{\mathpalette\underleftarrow@AmS}

\def\underleftarrow@AmS#1#2{\vtop{\halign{$##$\cr
     \hfil#1#2\hfil\cr
     \noalign{\kern -1pt\nointerlineskip}
     #1{\leftarrow}\mkern-6mu\cleaders\hbox{$#1\mkern-2mu{-}\mkern-2mu$}\hfill
      \mkern-6mu{-}\cr}}}

\def\underleftrightarrow{\mathpalette\underleftrightarrow@AmS}

\def\underleftrightarrow@AmS#1#2{\vtop{\halign{$##$\cr
      \hfil#1#2\hfil\cr
    \noalign{\kern -1pt\nointerlineskip}
     #1{\leftarrow}\mkern-6mu\cleaders\hbox{$#1\mkern-2mu{-}\mkern-2mu$}\hfill
       \mkern-6mu{\to}\cr}}}

% DOTS

\def\dotsc{\mathinner{\ldotp\ldotp\ldotp}}
\def\dotsi{\mathinner{\cdotp\cdotp\cdotp}}
\def\dotsj{\mathinner{\ldotp\ldotp\ldotp}}
\def\dotsb{\mathinner{\cdotp\cdotp\cdotp}}

\def\binary@AmS#1{{\thinmuskip 0mu \medmuskip 1mu \thickmuskip 1mu    % NOTE 74
      \setbox0=\hbox{$#1{}{}{}{}{}{}{}{}{}$}\setbox1=\hbox
       {${}#1{}{}{}{}{}{}{}{}{}$}\ifdim\wd1>\wd0\gdef\binary@@AmS{T}\else
       \gdef\binary@@AmS{F}\fi}}

\def\dots{\relax@AmS\ifmmode\def\dots@AmS{\mdots@AmS}\else
    \def\dots@AmS{\tdots@AmS}\fi\dots@AmS}

\def\mdots@AmS{\futurelet\tok@AmS\mdots@@AmS}

\def\mdots@@AmS{\def\thedots@AmS{\dotsj}%
  \ifx\tok@AmS\bgroup\else
  \ifx\tok@AmS\egroup\else
  \ifx\tok@AmS$\else
  \iffalse{\fi  \ifx\tok@AmS\\ \iffalse}\fi\else                      % NOTE 75
  \iffalse{\fi \ifx\tok@AmS&  \iffalse}\fi\else
  \ifx\tok@AmS\left\else
  \ifx\tok@AmS\right\else
  \ifx\tok@AmS,\def\thedots@AmS{\dotsc}\else
  \inttest@AmS\tok@AmS\if T\intflag@AmS\def\thedots@AmS{\dotsi}\else
  \binary@AmS\tok@AmS\if T\binary@@AmS\def\thedots@AmS{\dotsb}\else
   \def\thedots@AmS{\dotsj}\fi\fi\fi\fi\fi\fi\fi\fi\fi\fi\thedots@AmS}

\def\tdots@AmS{\unskip\ \tdots@@AmS}

\def\tdots@@AmS{\futurelet\tok@AmS\tdots@@@AmS}

\def\tdots@@@AmS{$\ldots\,
   \ifx\tok@AmS,$\else
   \ifx\tok@AmS.\,$\else
   \ifx\tok@AmS;\,$\else
   \ifx\tok@AmS:\,$\else
   \ifx\tok@AmS?\,$\else
   \ifx\tok@AmS!\,$\else
   $\ \fi\fi\fi\fi\fi\fi}

% SET NOTATION

\def\leftset#1\mid#2\rightset{\hbox{$\displaystyle
\left\{\,#1\vphantom{#1#2}\;\right|\;\left.
    #2\vphantom{#1#2}\,\right\}\offspace@AmS$}}

% ACCENT SYMBOLS

\def\dotii#1{{\mathop{#1}\limits^{\vbox to -1.4pt{\kern-2pt
   \hbox{\tenrm..}\vss}}}}
\def\dotiii#1{{\mathop{#1}\limits^{\vbox to -1.4pt{\kern-2pt
   \hbox{\tenrm...}\vss}}}}
\def\dotiv#1{{\mathop{#1}\limits^{\vbox to -1.4pt{\kern-2pt
   \hbox{\tenrm....}\vss}}}}

\def\hatsymbol{{\mathchoice{\null}{\null}{\,\,\hbox{\lower 10pt\hbox
    {$\widehat{\null}$}}}{\,\hbox{\lower 20pt\hbox
       {$\hat{\null}$}}}}}

% OVERSET AND OVERBRACE

\def\overset#1\to#2{{\mathop{#2}^{#1}}}

\def\underset#1\to#2{{\mathop{#2}_{#1}}}

\def\oversetbrace#1\to#2{{\overbrace{#2}^{#1}}}
\def\undersetbrace#1\to#2{{\underbrace{#2}_{#1}}}

% ROOTS

\def\theuproot{0 pt}

\def\therightroot{0mu}

\def\r@@t#1#2{\setbox\z@\hbox{$\m@th#1\sqrt{#2}$}%
  \dimen@\ht\z@ \advance\dimen@-\dp\z@ \advance\dimen@\theuproot
  \mskip5mu\raise.6\dimen@\copy\rootbox \mskip-10mu \mskip\therightroot
    \box\z@\gdef\theuproot{0 pt}\gdef\therightroot{0mu}}              % NOTE 76

%  BOXED

\def\boxed#1{\setbox0=\hbox{$\displaystyle{#1}$}\hbox{\lower.4pt\hbox{\lower
   3pt\hbox{\lower 1\dp0\hbox{\vbox{\hrule height .4pt \hbox{\vrule width
   .4pt \hskip 3pt\vbox{\vskip 3pt\box0\vskip3pt}\hskip 3pt \vrule width
      .4pt}\hrule height .4pt}}}}}}

%  FORMATTING MACROS COMMON TO ALL STYLES

\newif\ifretry@AmS
\def\y@AmS{y } \def\y@@AmS{Y } \def\n@AmS{n } \def\n@@AmS{N }
\def\ask@AmS{\message
  {Do you want output? (y or n, follow answer by return) }\loop
   \read-1 to\answer@AmS
  \ifx\answer@AmS\y@AmS\retry@AmSfalse\outputon
   \else\ifx\answer@AmS\y@@AmS\retry@AmSfalse\outputon
    \else\ifx\answer@AmS\n@AmS\retry@AmSfalse\outputoff
     \else\ifx\answer@AmS\n@@AmS\retry@AmSfalse\outputoff
      \else \retry@AmStrue\fi\fi\fi\fi
  \ifretry@AmS\message{Type y or n, follow answer by return: }\repeat}

\def\outputoff{\global\output{\setbox0=\box255 \deadcycles=0}}

\def\outputon{\global\output{\output@AmS}}

\catcode`\@=13

% NOTE 1, NOTE 2, ... REFER TO NOTES IN THE FILE AMSPPT.DOC

\catcode`\@=11

%AJS3
% This is the OLD version of AMSPPT.STY
% modified so as to use CM rather than AM fonts.

%  PARAMETERS DIFFERENT THAN IN PLAIN

\normallineskiplimit=1pt
\parindent 10pt
\hsize 26pc
\vsize 42pc

% EXTRA FONTS NEEDED

\font\eightrm=cmr8
\font\sixrm=cmr6
\font\eighti=cmmi8 \skewchar\eighti='177
\font\sixi=cmmi6 \skewchar\sixi='177
\font\eightsy=cmsy8 \skewchar\eightsy='60
\font\sixsy=cmsy6 \skewchar\sixsy='60
\font\eightbf=cmbx8
\font\sixbf=cmbx6
\font\eightsl=cmsl8
\font\eightit=cmti8
\font\tensmc=cmcsc10

% added by AJS3 for ninepoint

\font\ninerm=cmr9
\font\ninei=cmmi9 \skewchar\ninei='177
\font\ninesy=cmsy9 \skewchar\ninesy='60
\font\ninebf=cmbx9
\font\ninesl=cmsl9
\font\nineit=cmti9

% TWO DIFFERENT POINT SIZES

\def\tenpoint{\def\pointsize@AmS{t}\normalbaselineskip=12pt            % NOTE 1
 \abovedisplayskip 12pt plus 3pt minus 9pt
 \belowdisplayskip 12pt plus 3pt minus 9pt
 \abovedisplayshortskip 0pt plus 3pt
 \belowdisplayshortskip 7pt plus 3pt minus 4pt
 \def\rm{\fam0\tenrm}%
 \def\it{\fam\itfam\tenit}%
 \def\sl{\fam\slfam\tensl}%
 \def\bf{\fam\bffam\tenbf}%
 \def\smc{\tensmc}%
 \def\mit{\fam 1}%
 \def\cal{\fam 2}%
 \textfont0=\tenrm   \scriptfont0=\sevenrm   \scriptscriptfont0=\fiverm
 \textfont1=\teni    \scriptfont1=\seveni    \scriptscriptfont1=\fivei
 \textfont2=\tensy   \scriptfont2=\sevensy   \scriptscriptfont2=\fivesy
 \textfont3=\tenex   \scriptfont3=\tenex     \scriptscriptfont3=\tenex
 \textfont\itfam=\tenit
 \textfont\slfam=\tensl
 \textfont\bffam=\tenbf \scriptfont\bffam=\sevenbf
   \scriptscriptfont\bffam=\fivebf
\normalbaselines\rm}

\def\eightpoint{\def\pointsize@AmS{8}\normalbaselineskip=10pt
 \abovedisplayskip 10pt plus 2.4pt minus 7.2pt
 \belowdisplayskip 10pt plus 2.4pt minus 7.2pt
 \abovedisplayshortskip 0pt plus 2.4pt
 \belowdisplayshortskip 5.6pt plus 2.4pt minus 3.2pt
 \def\rm{\fam0\eightrm}%
 \def\it{\fam\itfam\eightit}%
 \def\sl{\fam\slfam\eightsl}%
 \def\bf{\fam\bffam\eightbf}%
 \def\mit{\fam 1}%
 \def\cal{\fam 2}%
 \textfont0=\eightrm   \scriptfont0=\sixrm   \scriptscriptfont0=\fiverm
 \textfont1=\eighti    \scriptfont1=\sixi    \scriptscriptfont1=\fivei
 \textfont2=\eightsy   \scriptfont2=\sixsy   \scriptscriptfont2=\fivesy
 \textfont3=\tenex   \scriptfont3=\tenex     \scriptscriptfont3=\tenex
 \textfont\itfam=\eightit
 \textfont\slfam=\eightsl
 \textfont\bffam=\eightbf \scriptfont\bffam=\sixbf
   \scriptscriptfont\bffam=\fivebf
\normalbaselines\rm}

% added by AJS3

\def\ninepoint{\def\pointsize@AmS{9}\normalbaselineskip=11pt
 \abovedisplayskip 11pt plus 2.7pt minus 8.1pt
 \belowdisplayskip 11pt plus 2.7pt minus 8.1pt
 \abovedisplayshortskip 0pt plus 2.7pt
 \belowdisplayshortskip 6.3pt plus 2.7pt minus 3.6pt
 \def\rm{\fam0\ninerm}%
 \def\it{\fam\itfam\nineit}%
 \def\sl{\fam\slfam\ninesl}%
 \def\bf{\fam\bffam\ninebf}%
 \def\mit{\fam 1}%
 \def\cal{\fam 2}%
 \textfont0=\ninerm   \scriptfont0=\sevenrm   \scriptscriptfont0=\fiverm
 \textfont1=\ninei    \scriptfont1=\seveni    \scriptscriptfont1=\fivei
 \textfont2=\ninesy   \scriptfont2=\sevensy   \scriptscriptfont2=\fivesy
 \textfont3=\tenex   \scriptfont3=\tenex     \scriptscriptfont3=\tenex
 \textfont\itfam=\nineit
 \textfont\slfam=\ninesl
 \textfont\bffam=\ninebf \scriptfont\bffam=\sevenbf
   \scriptscriptfont\bffam=\fivebf
\normalbaselines\rm}

%  GENERAL FOOTNOTES                                                   % NOTE 2

\newcount\footmarkcount@AmS
\footmarkcount@AmS=0
\newcount\foottextcount@AmS
\foottextcount@AmS=0

\def\footnotemark{\unskip\futurelet\tok@AmS\footnotemark@AmS}
\def\footnotemark@AmS{\ifx [\tok@AmS \def\next@AmS{\footnotemark@@AmS}\else
   \def\next@AmS{\footnotemark@@@AmS}\fi\next@AmS}
\def\footnotemark@@AmS[#1]{{#1}}
\def\footnotemark@@@AmS{\global\advance\footmarkcount@AmS by 1
 \xdef\thefootmarkcount@AmS{\the\footmarkcount@AmS}$^{\thefootmarkcount@AmS}$}

\def\makefootnote@AmS#1#2{\insert\footins{\interlinepenalty100
   \eightpoint
  \splittopskip=6.8pt
  \splitmaxdepth=2.8pt
   \floatingpenalty=20000
   \leftskip = 0pt  \rightskip = 0pt
    \noindent {#1}\footstrut{\ignorespaces#2\unskip}\topsmash{\strut}}}

\def\footnotetext{\futurelet\tok@AmS\footnotetext@}
\def\footnotetext@{\ifx [\tok@AmS \def\next@AmS{\footnotetext@@AmS}\else
  \def\next@AmS{\footnotetext@@@AmS}\fi\next@AmS}
\def\footnotetext@@AmS[#1]#2{\makefootnote@AmS{#1}{#2}}
\def\footnotetext@@@AmS#1{\global\advance\foottextcount@AmS by 1
  \xdef\thefoottextcount@AmS{\the\foottextcount@AmS}%
\makefootnote@AmS{$^{\thefoottextcount@AmS}$}{#1}}

\def\footnote{\unskip\futurelet\tok@AmS\footnote@AmS}
\def\footnote@AmS{\ifx [\tok@AmS \def\next@AmS{\footnote@@AmS}\else
   \def\next@AmS{\footnote@@@AmS}\fi\next@AmS}
\def\footnote@@AmS[#1]#2{{\edef\sf{\the\spacefactor}%
  {#1}\makefootnote@AmS{#1}{#2}\spacefactor=\sf}}
\def\footnote@@@AmS#1{\ifnum\footmarkcount@AmS=\foottextcount@AmS\else
 \errmessage{AmS-TeX warning: last footnote marker was \the\footmarkcount@AmS,
   last footnote was
   \the\foottextcount@AmS}\footmarkcount@AmS=\foottextcount@AmS\fi
   {\edef\sf{\the\spacefactor}\footnotemark@@@AmS\footnotetext@@@AmS{#1}%
    \spacefactor=\sf}}

\def\adjustfootnotemark#1{\advance\footmarkcount@AmS by #1}           % NOTE 3
\def\adjustfootnote#1{\advance\foottextcount@AmS by #1}

%  TOP MATTER

\def\topmatter@AmS{F}                                                 % NOTE 4
\def\topmatter{\def\topmatter@AmS{T}}

\def\filhss@AmS{plus 1000pt}                                          % NOTE 5
\def\overlong{\def\filhss@AmS{plus 1000pt minus1000pt}}

\newbox\titlebox@AmS

\setbox\titlebox@AmS=\vbox{}                                          % NOTE 6

\def\title#1\endtitle{{\let\\=\cr                                     % NOTE 7
  \global\setbox\titlebox@AmS=\vbox{\tabskip0pt\filhss@AmS
  \halign to \hsize
    {\tenpoint\bf\hfil\ignorespaces##\unskip\hfil\cr#1\cr}}}\def     % NOTE 7.1
     \filhss@AmSs{plus 1000pt}}

\def\isauthor@AmS{F}                                                % NOTE 8
\newbox\authorbox@AmS

\def\author#1\endauthor{\gdef\isauthor@AmS{T}{\let\\=\cr
 \global\setbox\authorbox@AmS=\vbox{\tabskip0pt
 \filhss@AmS\halign to \hsize
   {\tenpoint\smc\hfil\ignorespaces##\unskip\hfil\cr#1\cr}}}\def
      \filhss@AmS{plus 1000pt}}

%  FOR UPPERCASING TITLE AND AUTHOR

\def\uctext@AmS#1{\uppercase@AmS#1\gdef                           % NOTE 9
       \uppercase@@AmS{}${\hskip-2\mathsurround}$}
\def\uppercase@AmS#1$#2${\gdef\uppercase@@AmS{\uppercase@AmS}\uppercase
    {#1}${#2}$\uppercase@@AmS}

\newcount\Notes@AmS                                             % NOTE 10

\def\sfootnote@AmS{\unskip\futurelet\tok@AmS\sfootnote@@AmS}
\def\sfootnote@@AmS{\ifx [\tok@AmS \def\next@AmS{\sfootnote@@@AmS}\else
    \def\next@AmS{\sfootnote@@@@AmS}\fi\next@AmS}
\def\sfootnote@@@AmS[#1]#2{\global\toks@{#2}\advance\Notes@AmS by 1
  \expandafter\xdef\csname Note\romannumeral\Notes@AmS @AmS\endcsname
   {\the\toks@}}
\def\sfootnote@@@@AmS#1{\global\toks@{#1}\global\advance\Notes@AmS by 1
  \expandafter\xdef\csname Note\romannumeral\Notes@AmS @AmS\endcsname
  {\the\toks@}}

\def\Sfootnote@AmS{\unskip\futurelet\tok@AmS\Sfootnote@@AmS}
\def\Sfootnote@@AmS{\ifx [\tok@AmS \def\next@AmS{\Sfootnote@@@AmS}\else
    \def\next@AmS{\Sfootnote@@@@AmS}\fi\next@AmS}
\def\Sfootnote@@@AmS[#1]#2{{#1}\advance\Notes@AmS by 1
  {\edef\sf{\the\spacefactor}\makefootnote@AmS{#1}{\csname
     Note\romannumeral\Notes@AmS @AmS\endcsname}\spacefactor=\sf}}
\def\Sfootnote@@@@AmS#1{\ifnum\footmarkcount@AmS=\foottextcount@AmS\else
 \errmessage{AmS-TeX warning: last footnote marker was \the\footmarkcount@AmS,
  last footnote was
   \the\foottextcount@AmS}\footmarkcount@AmS=\foottextcount@AmS\fi
 {\edef\sf{\the\spacefactor}\footnotemark@@@AmS \global\advance\Notes@AmS by 1
    \footnotetext@@@AmS{\csname
      Note\romannumeral\Notes@AmS @AmS\endcsname}\spacefactor=\sf}}

\def\TITLE#1\endTITLE                                           % NOTE 11
{{\Notes@AmS=0 \let\\=\cr\let\footnote=\sfootnote@AmS
   \setbox0=\vbox{\tabskip\centering
  \halign to \hsize{\tenpoint\bf\ignorespaces##\unskip\cr#1\cr}}
 \Notes@AmS=0   \let\footnote=\Sfootnote@AmS
   \global\setbox\titlebox@AmS=\vbox{\tabskip0pt\filhss@AmS
\halign to \hsize{\tenpoint\bf\hfil
 \uctext@AmS{\ignorespaces##\unskip}\hfil\cr
          #1\cr}}}\def\filhss@AmS{plus 1000pt}}

\def\AUTHOR#1\endAUTHOR{\gdef\isauthor@AmS{T}{\Notes@AmS=0 \let\\=\cr
   \let\footnote=\sfootnote@AmS
 \setbox0 =\vbox{\tabskip\centering\halign to \hsize{\tenpoint\smc
   \ignorespaces##\unskip\cr#1\cr}}\Notes@AmS=0
   \let\footnote=\Sfootnote@AmS
  \global\setbox\authorbox@AmS=\vbox{\tabskip0pt\filhss@AmS\halign
  to \hsize{\tenpoint\smc\hfil\uppercase{\ignorespaces
     ##\unskip}\hfil\cr#1\cr}}}\def\filhss@AmS{plus 1000pt}}

% OTHER STUFF BEFORE \endtopmatter

\newcount\language@AmS                                            % NOTE 12
\language@AmS=0
\def\german{\language@AmS=1}

\def\abstractword@AmS{\ifcase \language@AmS ABSTRACT\or ZUSAMMENFASSUNG\fi}
\def\logoword@AmS{\ifcase \language@AmS Typeset by \fi}
\def\subjclassword@AmS{\ifcase \language@AmS
     1980 Mathematics subject classifications \fi}
\def\keywordsword@AmS{\ifcase \language@AmS  Keywords and phrases\fi}
\def\Referenceword@AmS{\ifcase \language@AmS References\fi}

\def\isaffil@AmS{F}
\newbox\affilbox@AmS
\def\affil{\gdef\isaffil@AmS{T}\bgroup\let\\=\cr
   \global\setbox\affilbox@AmS
     =\vbox\bgroup\tabskip0pt\filhss@AmS
 \halign to \hsize\bgroup\tenpoint\hfil\ignorespaces##\unskip\hfil\cr}

\def\endaffil{\cr\egroup\egroup\egroup\def\filhss@AmS{plus 1000pt}}

\newcount\addresscount@AmS                                         % NOTE 13
\addresscount@AmS=0

\def\address#1{\global\advance\addresscount@AmS by 1
  \expandafter\gdef\csname address\romannumeral\addresscount@AmS\endcsname
   {\noindent\eightpoint\ignorespaces#1\par}}

\def\isdate@AmS{F}                                                 % NOTE 14
\def\date#1{\gdef\isdate@AmS{T}\gdef\date@AmS{\tenpoint\ignorespaces#1\unskip}}

\def\isthanks@AmS{F}
\def\thanks#1{\gdef\isthanks@AmS{T}\gdef\thanks@AmS{\eightpoint\ignorespaces
       #1\unskip}}

\def\keywords@AmS{}                                                % NOTE 15
\def\keywords#1{\def\keywords@AmS{\noindent \eightpoint \it
\keywordsword@AmS .\enspace \rm\ignorespaces#1\par}}

\def\subjclass@AmS{}
\def\subjclass#1{\def\subjclass@AmS{\noindent \eightpoint\it
\subjclassword@AmS
(Amer.\ Math.\ Soc.)\/\rm: \ignorespaces#1\par}}

\def\isabstract@AmS{F}
\long\def\abstract#1{\gdef\isabstract@AmS{T}\long\gdef\abstract@AmS
   {\eightpoint \abstractword@AmS\period\ignorespaces #1\par}}        % NOTE 16

% ALLOW FOR SPECIAL THINGS BEFORE PARTS OF THE TOPMATTER              % NOTE 17

\def\pretitle{}
\def\preauthor{}
\def\preaffil{}
\def\predate{}
\def\preabstract{}
\def\prepaper{}

% \endtopmatter PUTS ALL THE PRELIMINARY STUFF ON THE FIRST PAGE IN PLACE

\def\endtopmatter{\if F\topmatter@AmS \errmessage{AmS-TeX warning: You
    forgot the \string\topmatter, but I forgive you.}\fi
\hrule height 0pt \vskip -\topskip                                   % NOTE 18
   \pretitle
   \vskip 24pt plus 12pt minus 12pt
   \unvbox\titlebox@AmS                                              % NOTE 19
   \preauthor
   \if T\isauthor@AmS \vskip 12pt plus 6pt minus 3pt
       \unvbox\authorbox@AmS \else\fi
    \preaffil
   \if T\isaffil@AmS \vskip 10pt plus 5pt minus 2pt
       \unvbox\affilbox@AmS\else\fi
  \predate
   \if T\isdate@AmS \vskip 6pt plus 2pt minus 1pt
  \hbox to \hsize{\hfil\date@AmS\hfil}\else\fi
    \preabstract
\if T\isthanks@AmS
  \makefootnote@AmS{}{\thanks@AmS}\else\fi
   \if T\isabstract@AmS \vskip 15pt plus 12pt minus 12pt
 {\leftskip=16pt\rightskip=16pt
  \noindent \abstract@AmS}\else\fi
   \prepaper
     \vskip 18pt plus 12pt minus 6pt \tenpoint}

% \enddocument PUTS ALL THE PRELIMINARY STUFF THAT APPEARS AT THE END IN PLACE

\newcount\addresnum@AmS                                               % NOTE 20
\def\enddocument{\penalty10000 \sfcode`\.3000\vskip 12pt minus 6pt  % NOTE 20.1
\keywords@AmS                                                         % NOTE 21
\subjclass@AmS
\addresnum@AmS=0
  \loop\ifnum\addresnum@AmS<\addresscount@AmS\advance\addresnum@AmS by 1
  \csname address\romannumeral\addresnum@AmS\endcsname\repeat
\vfill\supereject\end}

% HEADINGS AND SUBHEADINGS

\newbox\headingbox@AmS
\outer\def\heading{\medbreak\bgroup\let\\=\cr
\global\setbox\headingbox@AmS=\vbox\bgroup\tabskip0pt\filhss@AmS      % NOTE 22
   \halign to \hsize\bgroup\tenpoint\smc\hfil\ignorespaces
            ##\unskip\hfil\cr}

\def\endheading{\cr\egroup\egroup\egroup\unvbox\headingbox@AmS
    \penalty10000 \def\filhss@AmS{plus 1000pt}\medskip}

%  PROCLAIM AND DEMO, ETC.

\outer\def\proclaim#1{\xdef\curfont@AmS{\the\font}\medbreak        % NOTE 23
  \noindent\smc\ignorespaces#1\unskip.\enspace\sl\ignorespaces}

\outer\def\proclaimnp#1{\xdef\curfont@AmS{\the\font}\medbreak      % NOTE 24
  \noindent\smc\ignorespaces#1\enspace\sl\ignorespaces}

\def\finishproclaim{\par\curfont@AmS\ifdim\lastskip<\medskipamount % NOTE 25
 \removelastskip \penalty 55\medskip\fi}

\outer\def\demo#1{\par\ifdim\lastskip<\smallskipamount
  \removelastskip\smallskip\fi\noindent{\smc\ignorespaces#1\unskip:}\enspace
     \ignorespaces}

\outer\def\demonp#1{\ifdim\lastskip<\smallskipamount
  \removelastskip\smallskip\fi\noindent{\smc#1}\enspace\ignorespaces}

\newif\ifrunin@AmS                                                    % NOTE 27
\runin@AmSfalse
\def\runin{\runin@AmStrue}
\def\conditions{\def\\##1:{\par\noindent                              % NOTE 28
   \hbox to 1.5\parindent{\hss\rm\ignorespaces##1\unskip}%
      \hskip .5\parindent \hangafter1\hangindent2\parindent\ignorespaces}%
    \def\firstcon@AmS##1:{\ifrunin@AmS
     {\rm\ignorespaces##1\unskip}\ \ignorespaces
  \else\par\ifdim\lastskip<\smallskipamount\removelastskip\penalty55
     \smallskip\fi
     \\##1:\fi}\firstcon@AmS}
\def\endconditions{\par\smallbreak\runin@AmSfalse}                    % NOTE 29

% STUFF FOR BIBLIOGRAPHY

\def\refto#1{\in@AmS,{#1}\if T\cresult@AmS\refto@AmS#1\end@AmS\else   % NOTE 30
    [{\bf#1}]\fi}
\def\refto@AmS#1,#2\end@AmS{[{\bf#1},#2]}

\def\Refs{\bigbreak\hbox to \hsize{\hfil\tenpoint
    \smc \Referenceword@AmS\hfil}\penalty 10000
      \bigskip\eightpoint\sfcode`.=1000 }                             % NOTE 31

\newbox\nobox@AmS        \newbox\keybox@AmS        \newbox\bybox@AmS  % NOTE 32
\newbox\bysamebox@AmS    \newbox\paperbox@AmS      \newbox\paperinfobox@AmS
\newbox\jourbox@AmS      \newbox\volbox@AmS        \newbox\issuebox@AmS
\newbox\yrbox@AmS        \newbox\pagesbox@AmS      \newbox\bookbox@AmS
\newbox\bookinfobox@AmS  \newbox\publbox@AmS       \newbox\publaddrbox@AmS
\newbox\finalinfobox@AmS

\def\refset@AmS#1{\expandafter\gdef\csname is\expandafter\eat@AmS     % NOTE 33
  \string#1@AmS\endcsname{F}\expandafter
  \setbox\csname \expandafter\eat@AmS\string#1box@AmS\endcsname=\null}

\def\ref@AmS{\refset@AmS\no \refset@AmS\key \refset@AmS\by            % NOTE 34
\gdef\isbysame@AmS{F}%                                              % NOTE 35.1
 \refset@AmS\paper
  \refset@AmS\paperinfo \refset@AmS\jour \refset@AmS\vol
  \refset@AmS\issue \refset@AmS\yr
  \gdef\istoappear@AmS{F}%                                          % NOTE 35.2
  \refset@AmS\pages
  \gdef\ispage@AmS{F}%                                              % NOTE 35.3
  \refset@AmS\book
  \gdef\isinbook@AmS{F}%                                            % NOTE 35.4
  \refset@AmS\bookinfo \refset@AmS\publ
  \refset@AmS\publaddr \refset@AmS\finalinfo \bgroup
     \ignorespaces}                                                   % NOTE 36

\def\ref{\noindent\hangindent 20pt \hangafter 1 \def\refi@AmS{T}%     % NOTE 37
  \def\refl@AmS{F}\def\\{\egroup\endref@AmS\gdef\refi@AmS{F}\ref@AmS}\ref@AmS}

\def\refdef@AmS#1#2{\def#1{\egroup\expandafter                        % NOTE 38
  \gdef\csname is\expandafter\eat@AmS
  \string#1@AmS\endcsname{T}\expandafter\setbox
   \csname \expandafter\eat@AmS\string#1box@AmS\endcsname=\hbox\bgroup#2}}

\refdef@AmS\no{} \refdef@AmS\key{} \refdef@AmS\by{}
\def\bysame{\egroup\gdef\isbysame@AmS{T}\bgroup}                    % NOTE 39.1
\refdef@AmS\paper\it
\refdef@AmS\paperinfo{} \refdef@AmS\jour{} \refdef@AmS\vol\bf
\refdef@AmS\issue{} \refdef@AmS\yr{}
\def\toappear{\egroup\gdef\istoappear@AmS{T}\bgroup}                % NOTE 39.2
\refdef@AmS\pages{}
\def\page{\egroup\gdef\ispage@AmS{T}\setbox
                 \pagesbox@AmS=\hbox\bgroup}                        % NOTE 39.3
\refdef@AmS\book{}
\def\inbook{\egroup\gdef\isinbook@AmS{T}\setbox
                               \bookbox@AmS=\hbox\bgroup}           % NOTE 39.4
\refdef@AmS\bookinfo{} \refdef@AmS\publ{}
\refdef@AmS\publaddr{}
\refdef@AmS\finalinfo{}

\def\setpunct@AmS{\def\prepunct@AmS{, }}                              % NOTE 40
\def\ppunbox@AmS#1{\prepunct@AmS\unhbox#1\unskip}                     % NOTE 41

\def\endref@AmS{\def\prepunct@AmS{}%                                  % NOTE 42
\if T\refi@AmS                                                      % NOTE 43.1
  \if F\isno@AmS\hbox to 10pt{}\else                                % NOTE 43.2
     \hbox to 20pt{\hss\unhbox\nobox@AmS\unskip. }\fi               % NOTE 43.3
  \if T\iskey@AmS \unhbox\keybox@AmS\unskip\ \fi                    % NOTE 43.4
  \if T\isby@AmS  \hbox{\unhcopy\bybox@AmS\unskip}\setpunct@AmS     % NOTE 43.5
         \setbox\bysamebox@AmS=\hbox{\unhcopy\bybox@AmS\unskip}\fi  % NOTE 43.6
  \if T\isbysame@AmS                                                % NOTE 43.7
   \hbox to \wd\bysamebox@AmS{\leaders\hrule\hfill}\setpunct@AmS\fi
 \fi                                                                % NOTE 43.8
  \if T\ispaper@AmS\ppunbox@AmS\paperbox@AmS\setpunct@AmS\fi          % NOTE 44
  \if T\ispaperinfo@AmS\ppunbox@AmS\paperinfobox@AmS\setpunct@AmS\fi  % NOTE 45
  \if T\isjour@AmS\ppunbox@AmS\jourbox@AmS\setpunct@AmS               % NOTE 46
     \if T\isvol@AmS \ \unhbox\volbox@AmS\unskip\setpunct@AmS\fi    % NOTE 46.1
     \if T\isissue@AmS \ \unhbox\issuebox@AmS\unskip\setpunct@AmS\fi% NOTE 46.2
     \if T\isyr@AmS \ (\unhbox\yrbox@AmS\unskip)\setpunct@AmS\fi    % NOTE 46.3
     \if T\istoappear@AmS \ (to appear)\setpunct@AmS\fi             % NOTE 46.4
     \if T\ispages@AmS \ppunbox@AmS\pagesbox@AmS\setpunct@AmS\fi    % NOTE 46.5
     \if T\ispage@AmS                                               % NOTE 46.6
           \prepunct@AmS p.\ \unhbox\pagesbox@AmS\unskip\setpunct@AmS\fi
     \fi                                                            % NOTE 46.7
  \if T\isbook@AmS \prepunct@AmS                                      % NOTE 47
                     ``\unhbox\bookbox@AmS\unskip''\setpunct@AmS\fi
  \if T\isinbook@AmS \prepunct@AmS                                    % NOTE 48
    \unskip\ in ``\unhbox\bookbox@AmS\unskip''\setpunct@AmS
       \gdef\isbook@AmS{T}\fi
  \if T\isbookinfo@AmS \ppunbox@AmS\bookinfobox@AmS\setpunct@AmS\fi   % NOTE 49
  \if T\ispubl@AmS \ppunbox@AmS\publbox@AmS\setpunct@AmS\fi           % NOTE 50
  \if T\ispubladdr@AmS \ppunbox@AmS\publaddrbox@AmS\setpunct@AmS\fi   % NOTE 51
 \if T\isbook@AmS                                                     % NOTE 52
  \if T\isyr@AmS \prepunct@AmS \unhbox\yrbox@AmS\unskip             % NOTE 52.1
              \setpunct@AmS\fi
  \if T\istoappear@AmS \ (to appear)\setpunct@AmS\fi                % NOTE 52.2
  \if T\ispages@AmS                                                 % NOTE 52.3
    \prepunct@AmS pp.\ \unhbox\pagesbox@AmS\unskip\setpunct@AmS\fi
  \if T\ispage@AmS                                                  % NOTE 52.4
    \prepunct@AmS p.\ \unhbox\pagesbox@AmS\unskip\setpunct@AmS\fi
 \fi
  \if T\isfinalinfo@AmS \period\unhbox\finalinfobox@AmS\else          % NOTE 53
    \if T\refl@AmS .\else ; \fi\fi}

\def\endref{\egroup\gdef\refl@AmS{T}\endref@AmS\par}

%  OUTPUT

\newif\ifguides@AmS
\guides@AmSfalse
\def\guidelines{\guides@AmStrue}
\def\noguidelines{\guides@AmSfalse}
\def\guidelinegap#1{\def\gwidth@AmS{#1}}
\def\gwidth@AmS{24pt}

\newif\iflogo@AmS
\def\nologo{\logo@AmSfalse}
\logo@AmStrue

\def\output@AmS{\ifnum\count0=1
 \shipout\vbox{\ifguides@AmS\hrule width \hsize \vskip\gwidth@AmS \fi
   \vbox to \vsize{\boxmaxdepth=\maxdepth\pagecontents}\baselineskip2pc
\iflogo@AmS \hbox to \hsize{\hfil\eightpoint \logoword@AmS\AmSTeX}\fi
     \ifguides@AmS \vskip\gwidth@AmS
\hrule width \hsize\fi}\vsize 44pc\else
   \shipout\vbox{\ifguides@AmS \hrule width \hsize \vskip\gwidth@AmS\fi
   \vbox to \vsize{\boxmaxdepth=\maxdepth\pagecontents}\baselineskip2pc\hbox to
  \hsize{\hfil \tenpoint\number\count0\hfil}\ifguides@AmS
    \vskip\gwidth@AmS\hrule width \hsize\fi}\fi\global\advance\count0 by 1
  \global\footmarkcount@AmS=0 \global\foottextcount@AmS=0
 \ifnum\outputpenalty>-20000 \else\dosupereject\fi}

 % the \hrule is .4pt high

%  INITIAL STUFF

%AJS3: commented out asking if you want output
%\ask@AmS

\tenpoint

\catcode`\@=13

\output={\plainoutput}

\magnification=\magstep1
\baselineskip=16pt
\hoffset=-0.75truecm
\voffset=0.0truecm
\vsize=23.5truecm
\hsize=18.0truecm
\parskip=0.2cm
\parindent=1cm

\hfuzz=10pt

\def \bigbreak  {\goodbreak\bigskip}
\def \medbreak  {\goodbreak\medskip}
\def \smallbreak{\goodbreak\smallskip}
\def \header#1{\goodbreak\bigskip\centerline{\bf #1}\medskip\nobreak}
\def \subheader#1{\goodbreak\medskip\par\noindent{\bf #1}\smallskip\nobreak}

%===================================================================
%
%	POOR MAN'S BOLDFACE FROM TEXBOOK
%
\def\pmb#1{\setbox0=\hbox{#1}%
  \kern-.025em\copy0\kern-\wd0
  \kern.05em\copy0\kern-\wd0
  \kern-.025em\raise.0433em\box0 }
%
%===================================================================
%
%	Time and Date in typewriter font
%
\def\timedate{ {\tt
\count215=\time \divide\count215 by60  \number\count215
\multiply\count215 by-60 \advance \count215 by\time :\number\count215 \space
\number\day\space
\ifcase\month\or January\or February\or March\or April\or May\or June\or July
\or August\or September\or October\or November\or December\fi\space\number\year
}}
%
%===================================================================
%\def\pp{\par\parshape 2 0truecm 18truecm 1truecm 17truecm \noindent}
\def \etal {{\it et al.} }

%
%	\captpar PARAGRAPH SHAPE FOR FIGURE CAPTIONS
%
\def\captpar{\dimen0=\hsize
             \advance\dimen0 by -1.0truecm
             \par\parshape 1 0.5truecm \dimen0 \noindent}
\def\pp{\dimen0=\hsize
        \advance\dimen0 by -1truecm
        \par\parshape 2 0truecm \dimen0 1truecm \dimen0 \noindent}
%
%  \pp PARAGRAPH SHAPE IN WHICH THE FIRST LINE IS NOT INDENTED, BUT SUBSEQUENT
%      LINES ARE.
%

\def\sqr#1#2{{\vcenter{\hrule height.#2pt
              \hbox{\vrule width.#2pt height#1pt \kern#1pt \vrule width.#2pt}
              \hrule height.#2pt}}}

\def\mathrelfun#1#2{\lower3.6pt\vbox{\baselineskip0pt\lineskip.9pt
  \ialign{$\mathsurround=0pt#1\hfil##\hfil$\crcr#2\crcr\sim\crcr}}}

\def\rmK {{\rm K}}

\def\eV  {{\rm \hbox{e\kern-0.14em V}}}
\def\keV {{\rm \hbox{ke\kern-0.14em V}}}
\def\MeV {{\rm \hbox{Me\kern-0.14em V}}}
\def\GeV {{\rm \hbox{Ge\kern-0.14em V}}}

\font\FermiPPTfont=cmssbx10 scaled 1440
\font\FermiSmallfont=cmssq8 scaled 1200

\def\FNALpptheadnologo#1#2{
\null \vskip -1truein
\centerline{\hbox to 7.5truein {
\hskip 1.5cm
\vbox to 1in{\vfill
             \hbox{{\FermiPPTfont Fermi National Accelerator Laboratory}}
             \vfill}
\hfill
\vbox to 1in {\vfill \FermiSmallfont
              \hbox{#1}
              \hbox{#2}
              \vfill}
}}}%FNALpptheadnologo

\FNALpptheadnologo{FERMILAB-Conf-94/197-A}{July 1994}

\topmatter
\title
Cosmic Microwave Background Radiation Anisotropy
Induced by Cosmic Strings
\footnote{\tt Presented at Case Western Reserve CMBR Conference (April 22-24,
          1994)}
\footnote{\tt Presented at the {\sl International Conference on Unified
          Symmetry} (Coral Gables) (January, 1994)}
\endtitle
\author
B. Allen${}^\spadesuit$, R. R. Caldwell${}^\clubsuit$,
E. P. S. Shellard${}^\diamondsuit$,
A. Stebbins${}^\clubsuit$, S. Veeraraghavan${}^\heartsuit$
\endauthor

\affil
$\spadesuit$ Department of Physics, University of Wisconsin,
                                                         Milwaukee, WI 53201 \\
$\clubsuit$ NASA/Fermilab Astrophysics Center, FNAL, Box 500,
                                                            Batavia, IL 60510\\
$\diamondsuit$ DAMTP, Cambridge University, Cambridge, CB3 0EH, U.K.\\
$\heartsuit$Goddard Space Flight Center Greenbelt, MD 20771
\endaffil

\abstract{\ninepoint
We report on a current investigation of the anisotropy pattern induced by
cosmic strings on the cosmic microwave background radiation (MBR).  We have
numerically evolved a network of cosmic strings from a redshift of $Z = 100$ to
the present and calculated the anisotropies which they induce.  Based on a
limited number of realizations, we have compared the results of our simulations
with the observations of the COBE-DMR experiment. We have obtained a
preliminary estimate of the string mass-per-unit-length $\mu$ in the cosmic
string scenario.}
\endtopmatter

\header{1. Introduction}

Cosmic strings are tubes of topologically bound quanta, which may have formed
during a phase transition during the early universe, and which may contribute
to the formation of large scale structure. The properties of cosmic strings
have been well investigated, both analytically and numerically, since the early
80's [1-3].  Roughly, the following understanding of how cosmic strings behave
on the large and small scales has developed. On large scales, long, wiggly
strings sweep out wakes in the cosmological fluid [4-10], leaving overdense
regions which may grow to form galaxies and clusters. On small scales, small
loops are chopped off and ejected from the long strings.  These loops may then
contract and expand under their own tension, radiating gravitational waves
[11]. Processes on both the large and small scales contribute to the gross
features of the cosmic string scenario, and the observational properties of
cosmic strings.  It is important to note that the understanding of the
cosmological properties of cosmic strings continues to evolve.  The development
of cosmic string cosmology is not due to modification of the model; no
`adjustment' or `twisting' of parameters has occurred.  Rather, progress
derives from the continued sophistication of the analytical and numerical
techniques applied to the study of cosmic strings.

In the cosmic string scenario for the formation of large scale structure, the
cosmological perturbations in the density and microwave background radiation
(MBR) are induced by the gravitational fields of the cosmic strings.  Present
estimates of the only free parameter of the model, $\mu$, the
mass-per-unit-length, suggest $G\mu/c^2 \sim1-4\times 10^{-6}$ [11-17].  That
is, for $\mu$ within this approximate range of values, cosmic strings produce a
spectrum of density and MBR fluctuations which appear to be in rough agreement
with observations. Here we examine the large-scale MBR anisotropies produced by
strings more carefully than has been done previously.

We may make several preliminary observations regarding the properties of the
fluctuations induced in the MBR by cosmic strings.  It is well understood that
a single, moving, cosmic string produces a non-Gaussian, discontinuous
temperature shift on a field of photons passing by the string [18].  On large
angular scales on the celestial sphere, however, the cumulative effect of many
cosmic strings present between the surface of last scattering and the observer
will render less apparent the features of individual strings.  The effects of
individual strings should be most apparent on smaller angular scales [19-20].
Additional anisotropies will also arise from the perturbations induced in the
baryons and dark matter by the strings. Cosmic strings, as well as other
topological defects, produce density inhomogeneities; the spectrum of
anisotropies will exhibit the same type of ``doppler peak'' as do other models,
although the amplitude may be different [21].

Here we consider the large-scale anisotropy of the MBR, by computing the
large-scale temperature field on the celestial sphere from realizations of a
numerical simulation of cosmic strings.  Comparing predictions of the rms
temperature fluctuations with the COBE-DMR experiment [22], we may obtain an
estimate of $G\mu/c^2$.

\header{2. Method}

In this section we shall briefly discuss the method by which we simulate the
anisotropy pattern induced in the MBR by cosmic strings.  First, we shall
discuss the cosmic string simulation. Second, we shall discuss the procedure by
which cosmic string simulation data are translated into a map of the
temperature
field on the celestial sphere.

\subheader{Cosmic String Simulation}

The role of the cosmic string simulation in this research is to evolve a
network of cosmic strings which lie within the past light cone of an observer,
from some early time to the present. At each timestep in this interval
the cosmic string simulation code computes the position and velocity of
the cosmic string, modeled as cosmic string ``segments''.  From the positions,
velocities, and lengths of the segments we can obtain the stress-energy tensor,
$\Theta_{\mu\nu}$, of the strings, which is used to calculate the anisotropy as
described below.

The string simulation is a modified version of the Allen and Shellard code,
described in detail in [23]. The code has been modified to permit a model,
dust-dominated, $\Omega=1$, FRW universe to undergo a much longer period of
expansion than in previous work.  This is necessary in order to follow the
evolution of the cosmic strings over a sufficiently large redshift range.  We
have modified the code to maintain only a fixed number of cosmic
string segments per horizon length,
consistent with the angular resolution of the anisotropy calculation.  The
simulations used here have been
evolved from a redshift of $Z=100$ to the present,
which may be sufficient to include anisotropies produced on scales to which
COBE is sensitive.

The method by which a fixed level of resolution is maintained is fairly simple.
The basic idea is to correctly approximate the shape of the cosmic string on a
fixed comoving length scale. Thus, neighboring pairs of sufficiently short
cosmic string segments are combined by converting the adjacent pairs into a
single new segment.  The drawback is that this procedure does not specify the
direction of the momentum vector of the new segment; this procedure conserves
energy, but not momentum.  We may take the following precautions: the new
momentum vector must have length determined by the transverse velocity of the
new segment, lie in the plane perpendicular to the new segment, and be parallel
to the projection of the sum of the two original momentum vectors onto this
plane. Taken together, these conditions come as close as possible to satisfying
the requirements of energy and momentum conservation.  We have tested this
procedure by monitoring the energy and rms velocity along long strings, on
different length scales.  We find that the segment-joined, fixed-resolution
strings satisfactorily approximate the behavior of high resolution cosmic
strings.  We hope to soon make a quantitative statement regarding the degree to
which segment-joined strings approximate `realistic' cosmic strings.

\vskip 0.2in
\subheader{Computation of the Temperature Field}

Cosmic strings contribute only a small fraction of the cosmological density
parameter to the cosmological fluid.  We may estimate that
$\Omega_{\rm cosmic\,strings}\sim(2\pi / 3) A {G\mu/c^2} \sim 10^{-4}$ where
$A\sim 30$ is the average number of long strings present in a horizon volume.
Similarly, the cosmic string gravitational field will induce only small
perturbations in the surrounding matter and metric. The perturbations in the
matter will grow via gravitational instability, and may produce the structures
we observe today in the universe. The  metric perturbations induced directly by
the strings and indirectly via matter perturbations will remain weak. The
photons traveling through these weak fields, however, will gain or lose energy
via the Sachs-Wolfe effect [24], ultimately leading to the anisotropy observed
in the MBR.

One may use linear theory to express the anisotropy pattern in terms of the
stress-energy of the string:
$$({\delta T\over T})(\theta,\phi)_S
=\int d^4x G^{\mu\nu}_S\Theta_{\mu\nu}.\eqno(1)$$
Here, $S$ indicates the contribution by strings.  To compute the Green's
functions $G^{\mu\nu}_S$ one first calculates the metric perturbations in terms
of $\Theta_{\mu\nu}$, as in [25], and then inserts this integral solution
into the Sachs-Wolfe integral, which has been done in [26].  These Green's
functions have support on and within the past light cone of the observer.

Now, because the cosmic strings were created out of the cosmological fluid via
the Kibble mechanism [27] at some early time, the energy and momentum in the
strings was taken from the cosmological fluid.  There will remain an
anti-correlation between the energy-momentum of the strings and the
energy-momentum of the rest of the matter at all later times.  Although our
simulation evolves strings at times long after their formation, we include
matter perturbations in the initial data in order to ``compensate'' the energy
and momentum of the strings in the inital timestep of the simulation.
Thus, to the temperature field computed in
equation (1), we add a term which compensates the initial energy density of the
strings
$$({\Delta T\over T})(\theta,\phi)_I
=\int d^3x G^{00}_I \Theta_{00}.\eqno(2)$$
Here, $I$ indicates the initial compensation.  This term makes a significant
correction to the anisotropy at large angular scales. Compensation of the
momentum would only add a small further correction, and is neglected.  We
perform the integrals described in equations (1-2) using the discretized
representation of the cosmic string network of the simulation described above,
summing the contributions of each string segment at each time step to the
temperature pattern. The temperature pattern is itself discretized by
constructing a grid of 6144 pixels on the celestial sphere, each approximately
$3^\circ \times 3^\circ$ in size.  The 2-parameter ($\theta,\phi$),
4-dimensional integral of equations (1-2) may appear to be rather cumbersome.
Since the strings are, however, 1-dimensional objects, the effective
dimensionality of the integrals may be reduced.

We have carried out several preliminary tests of the code employed to compute
the temperature field. These tests compare the computed anisotropy pattern for
simple string configurations with known analytic results.  We find that the
results of the numerical simulation agree well with analytic calculations.
Further tests, as well, are currently being carried out.

\header{3. Results}

The output of the numerical simulation is the (discretized) temperature field
on the celestial sphere,
$${\Delta T \over T}(\theta,\phi).\eqno(3)$$
Here we present the anisotropy pattern as seen by two different observers
situated at different positions but viewing the same realization of the string
network. The patterns themselves are displayed in figures 1 and 2, using the
equal area, Hammer-Aitoff projection.  The monopole and dipole moments of
the anisotropy pattern have
been subtracted from the maps. No smoothing has been performed and the
pixels are easily visible.  The effect of this grid on the anisotropy pattern
smoothed over $10^\circ$, discussed below, should be negligible.  Furthermore,
a variety of systematic errors may be present, for which we have not corrected.
Therefore, we present {\it preliminary} results.

	The most striking features in these two maps are the adjacent hot and
cold spots in figure 1.  This is a result of a piece of string which is moving
extremely rapidly ($v=0.99c$) as it passes the observer's past light cone.  No
equally long and rapid string segment passes through the 2nd observer's past
light-cone.
We are currently investigating whether aspects of
the hot and cold spot are artifacts of our numerical computation.
Such a feature in our own sky might argue strongly for a scenario
like cosmic strings. The absence, however, would not necessarily argue against
a cosmic string scenario.
The statistics of such features need to be studied
in more detail.

	Using this temperature field we may compute the power spectrum of the
MBR fluctuations, defined by
$$C_l={1\over2l+1}\sum_{m=-l}^l|a_{lm}|^2 \qquad
{\Delta T\over T}(\theta,\phi)
                      =\sum_{l=0}^\infty\sum_{m=-l}^l a_{lm}Y_{lm}(\theta,\phi)
.\eqno(4)$$
In figure 3 we present the anisotropy spectrum,
$ [l(l+1) C_l/ 2\pi]^{1/2}/(G\mu/c^2)$, obtained from each
of the two maps.
Some difference between the two spectra is expected due to ``cosmic
variance''. The fall off at large $l$ is at least partly due to the
discretization of the temperature field on the celestial sphere. Based on this
small number of realizations, there also appears to be a fall off at low
multipole moments, although not as much as has been predicted [12,16].

The rms temperature anisotropy,  to be compared with the COBE-DMR signal after
subtracting the monopole and the dipole is
$$({\Delta T\over T})_{\rm r.m.s.}
=\Bigl( \sum_{l=2}^\infty {2l+1\over4\pi}\,W_l^2\,C_l \Bigr)^{1/2} \qquad
W_l^2 \equiv\exp[-l(l+1)/(13.5)^2]
\eqno(5)$$
where $W_l$ describes the smearing of the anisotropy by a Gaussian $10^\circ$
beam. Averaging the rms temperature for the two observers, we find
$$({\Delta T\over T})_{\rm r.m.s.}= 9 {G\mu\over c^2}.\eqno(6)$$
At present we can put no meaningful error bars on this quantity.  Comparing
this number with results obtained from the two-year COBE-DMR data [22],
$$(\Delta T)^{\rm DMR}_{r.m.s.} = (30.5 \pm 2.7)\times 10^{-6}\, \rmK
\qquad T=2.7 \,\rmK
\eqno(7)$$
we find that
$${G\mu\over c^2}= 1.3 \times 10^{-6}.\eqno(8)$$
Again, {\it this is only a preliminary estimate}.  Of course if there are other
contributions to the anisotropy from perturbations not induced by strings, one
should interpret this as an upper limit. This estimate is consistent with
previous estimates using the COBE data [12].

\header{4. Conclusion}

We have presented some preliminary results from a numerical simulation of the
MBR anisotropy induced by cosmic strings.  With the appropriate normalization
these anisotropies are consistent with the MBR anisotropies observed by the
COBE-DMR experiment.  Because of the sophistication of the string evolution and
the full-sky treatment, this work should ultimately provide the first
definitive normalization of the cosmic string scenario. It already broadly
confirms other estimates of MBR anisotropies from strings [12,20], and further
analysis of our techniques and results will allow us to better quantify the
uncertainties in this calculation.  In the near future this work will also
quantify the angular spectral index, the degree of non-gaussianity, and the
significance of cosmic variance.

With this overall normalization, cosmic strings will then have to be judged
against predictions for small-angle MBR anisotropies and through comparisons of
variant structure formation models with observation. Given our preliminary
value of $G\mu/c^2$, power spectrum analysis [13] already suggests that the
density fluctuations induced by cosmic strings are smaller in amplitude than
the observed galaxy number fluctuations.  Such biasing is desirable for
small-scale clustering but it may be harder to reconcile with reported
large-scale peculiar velocities [15].  However, string galaxy formation models
clearly deserve more quantitative investigation. Of course, if cosmic strings
are not responsible for structure formation, then MBR results currently provide
the most stringent upper bound on the energy scale for symmetry breaking phase
transitions which produce strings.

\vskip 0.2in
\noindent {\bf Acknowledgements} The work of RRC and AS is supported by the
NASA through  Grant No. NAGW-2381 (at Fermilab). The work of BA is supported by
the NSF through Grant No. PHY-91-05935 (at Milwaukee). The work of SV is
supported by the National Research Council.

\vfill\eject

\tenpoint

\vfill\eject

{\bf Figure 1} The anisotropy pattern induced by cosmic strings, for one
observer location in a cosmic string simulation, as described in the text. The
monopole and dipole moments have been subtracted.  Note the hot and cold patch
(adjacent yellow and blue spots) on the right side of the map, caused by an
ultra-relativistic piece of string. These patches saturate the temperature
scale, with a maximum amplitude of $100$. The temperature scale has been chosen
for comparison with figure 2.

{\bf Figure 2} The anisotropy pattern induced by cosmic strings, for a second
observer located at a different position but in the same string simulation.
The monopole and dipole moments have been subtracted.  To this observer there
are no hot or cold patches with comparable amplitude.  There is no saturation
of the temperature scale in this map.

{\bf Figure 3} The power spectrum of anisotropies, determined from the results
of the numerical simulation. The solid and dotted lines are for the first and
second observers (figures 1 and 2) respectively. The fall off in the spectrum
at large $l$ is at least partially due to the discretization of the temperature
field on the celestial sphere.  The amplitude of the spectrum is normalized by
the cosmic string mass-per-unit-length $\mu$.  Also shown is a model for the
power spectrum given in [12].

\end